\documentclass[
 reprint,
 amsmath,amssymb,
 aps,
 prd,
]{revtex4-2}

\usepackage{xcolor}
\usepackage[colorlinks=true, linkcolor=red, citecolor=blue, urlcolor=blue]{hyperref}
\usepackage{graphicx}
\usepackage{mathrsfs}
\usepackage{bm}
\usepackage{dcolumn}

\newcommand{\p}{p}
\newcommand{\pc}{p_{\rm th}}
\newcommand{\MBH}{M_{\rm PBH}}

\usepackage{ulem}

\newcommand{\Blue}[1]{{\color{blue} #1}}

\begin{document}

\title{Cosmological discrete self-similarity in primordial black hole formation}

\author{Luis E. Padilla}
\affiliation{Department of Physics, Rikkyo University, Tokyo, Japan}

\author{Tomohiro Harada}
\affiliation{Department of Physics, Rikkyo University, Tokyo, Japan}

\author{Ethan Milligan}
\affiliation{Astronomy Unit, Queen Mary University of London, Mile End Road, London, E1 4NS, UK}

\author{David Mulryne}
\affiliation{Astronomy Unit, Queen Mary University of London, Mile End Road, London, E1 4NS, UK}
\date{\today}

\begin{abstract}
We demonstrate that discrete self-similarity (DSS), originally discovered in the asymptotically flat collapse of a massless scalar field, survives in primordial black hole (PBH) formation within an expanding cosmological background. Using fully relativistic simulations of massless scalar-field collapse in a Friedmann-Lemaître-Robertson-Walker universe, we resolve the near-critical regime and find clear log-periodic oscillations in the PBH mass-scaling relation. {This result provides a proof of principle that PBHs may retain information about the critical behaviour {in their formation} realized in the early Universe: {PBH formation in scalar-field kination exhibits DSS,  while radiation-fluid collapse is characterized by continuous self-similarity.}} The resulting mass scaling imprints characteristic log-periodic modulations in an illustrative PBH mass spectrum, which may survive when PBH formation is dominated by a sufficiently narrow range of scales, with possible consequences for PBH phenomenology. {Our results show that PBHs may encode not only the amplitude of primordial perturbations, but also the nature of the matter sector and the dynamical state of the Universe at horizon re-entry.}
\end{abstract}

\maketitle

\textit{Introduction.}
Primordial black holes (PBHs) \citep{Zeldovich:1967lct,10.1093/mnras/152.1.75,10.1093/mnras/168.2.399, Byrnes:2025tji}
provide a sensitive probe of the early universe and physics at energy scales inaccessible to laboratories.
In many scenarios PBHs form from the collapse of large-amplitude perturbations shortly after horizon re-entry,
leading to a critical regime where the black-hole mass follows a universal scaling law
\begin{equation}\nonumber
\MBH \propto (\p-\pc)^{\gamma},
\label{eq:powerlaw_plain}
\end{equation}
where $p$ is a control parameter of the initial data, $\pc$ denotes its critical value separating dispersion from collapse,
and $\gamma$ is a universal critical exponent.
{Because the pre-BBN expansion history is poorly constrained, PBHs may probe not only the amplitude of primordial perturbations, but also the matter model and dynamical phase of the Universe at horizon re-entry\Blue{~\citep{Allahverdi:2020bys}.}}

Critical {phenomena in} gravitational collapse {were} discovered by Choptuik in {an asymptotically flat} spacetime for a massless scalar field \citep{PhysRevLett.70.9},
revealing both the scaling exponent $\gamma$ and a discretely self-similar (DSS) critical solution characterized by ``echoes''
periodic in logarithmic scale. In that setting, the mass scaling is more precisely described by \citep{Hod:1996az,PhysRevD.55.695}
\begin{equation}
\ln \MBH = \gamma \ln(\p-\pc) + f\left[\ln(\p-\pc)\right],
\label{eq:mass_scaling_with_wiggles}
\end{equation}
where $f$ is a periodic function encoding the DSS of the critical solution.
We denote by $P_{\ln}$ the period of $f$ in its argument $\ln(\p-\pc)$, so that
$f(x+P_{\ln})=f(x)$. In standard DSS collapse one may equivalently characterize self-similarity by an echoing period $\Delta$ in logarithmic self-similar time, with $P_{\ln}=\Delta/(2\gamma)$ \citep{Hod:1996az}.
{The above {log-periodic} feature is not seen in the critical behaviour caused by a continuously self-similar {(CSS)} solution observed in the perfect fluid system with a linear equation of state including radiation fluid.}
{For a timelike scalar-field gradient, a massless scalar admits a local perfect-fluid representation with effective equation of state $w=1$, which underlies our construction of super-horizon initial data below. Nevertheless, its nonlinear collapse dynamics is not equivalent to that of a perfect stiff fluid. This distinction is essential here: the DSS behavior studied below is a property of the massless scalar-field system and probes the field-theoretic nature of the dominant component, not only the background equation of state. See Brady {\it et al.}~\citep{Brady:2002iz}.}

{In the cosmological context relevant for PBH formation, critical collapse takes place within an expanding background that introduces a physical scale through the Hubble horizon, thereby breaking exact scale invariance.
This naturally raises the question of whether the DSS critical solution survives in such an environment.}
{Scalar-field kination is not an ad hoc alternative to radiation domination, but a well-studied post-inflationary phase arising in quintessential inflation, runaway or non-oscillatory inflation, moduli-dominated histories, and scalar-field reheating scenarios \citep{Spokoiny:1993kt,Joyce:1996cp,Peebles:1998qn,Gouttenoire:2021jhk}. PBH formation during kination has recently attracted interest because the non-standard expansion history can modify PBH abundances, the required small-scale primordial power, and induced gravitational-wave signals \citep{Bhattacharya:2019bvk,Dalianis:2021iig,Harigaya:2023wtn}, while nonlinear scalar inhomogeneities during kination have been studied with numerical relativity \citep{Cheng:2025hba}. This setting provides a proof of principle that PBHs can probe different forms of critical behaviour in the early Universe: radiation-fluid collapse exhibits CSS, whereas massless scalar-field kination exhibits DSS.}
In this Letter we demonstrate that DSS persists in cosmological massless scalar-field collapse, {and clarify how the near-critical self-similar solution emerges within an expanding background.}

\textit{Setup: cosmological scalar-field collapse.}
We perform fully relativistic numerical simulations of spherically symmetric collapse using a previously validated numerical code
based on the {extended} Misner--Sharp formalism \citep{Milligan:2025zbu,Padilla:2025bkv}.
The spacetime metric is written in comoving coordinates to a perfect fluid adapted to spherical symmetry,
\begin{equation}\nonumber
ds^2 = -e^{2\phi}dt^2 + e^{\lambda} dA^2 + R^2 d\Omega^2 ,
\end{equation}
 where $\phi{(t,A)}$ is the lapse function, $\lambda{(t,A)}$ encodes the radial metric component, $R(t,A)$ is the areal radius{, and $d\Omega^{2}$ is the metric on the unit sphere}.
This formulation leads to a coupled system of evolution and constraint equations for the metric and matter variables, which we solve numerically (see Supplemental Material~\cite{SupplementalMaterial} for details).

Initial conditions are constructed using the gradient-expansion formalism
\citep{Shibata_1999,Harada:2015yda},
which provides a systematic description of super-horizon perturbations in an expanding universe.
In this framework, all initial inhomogeneities are encoded in a single curvature perturbation function {
$K(A)$},
from which the metric and matter variables are consistently determined order by order in spatial gradients.
To leading non-trivial order, the dynamical variables admit expansions of the form
\begin{equation}\nonumber 
R = aA\left(1+\epsilon^2 \tilde R[K]\right), \qquad
\rho_\psi = \rho_{\rm b}\left(1+\epsilon^2 \tilde\rho_\psi[K]\right),
\end{equation}
where $\epsilon$ is the gradient-expansion parameter, $a$ is the scale factor,
and $\rho_{\rm b}$ denotes the background energy density.
This construction yields consistent initial data well outside the horizon,
which are subsequently evolved through horizon re-entry.

We consider a real, massless scalar field $\psi$. We use the gradient expansion and the initial curvature profile 
{
$K(A)$}
to construct the associated initial {scalar-field} energy density $\rho_\psi(t_0,A)$.
The scalar-field variables are {then} initialized as
\begin{equation}\nonumber 
\psi(t_0,A)= \partial_A\psi(t_0,A)=0, \qquad
\Pi(t_0,A)=\sqrt{2\,\rho_\psi(t_0,A)} ,
\end{equation}
where $\Pi\equiv e^{-\phi}\dot\psi$.
With this choice the scalar field is initially comoving with the coordinates,
$\nabla_\mu\psi\propto \delta^t_\mu$, 
so that the timelike-gradient condition required for the initial perfect-fluid
identification is satisfied by construction, while no restriction is imposed
on the subsequent dynamical evolution of the scalar field.

As initial conditions, we consider two families of curvature profiles $K(A)$: a Gaussian profile and a piecewise rational profile, given respectively by
\begin{equation}\nonumber
K_{\rm G}(A) = p\, e^{-(A / A_m)^2},
\end{equation}
and
\begin{equation}\nonumber
K_{\rm P}(A) = \frac{p[1-(3x^2-2x^3)]}{1+(A/A_m)^2},
\end{equation}
where
\begin{equation}\nonumber
    x(A) =
\begin{cases}
{ 0}, & A < A_m, \\
\frac{A-A_m}{4A_m},   & A_m < A < 5A_m, \\
{1},   & A > 5A_m.
\end{cases}
\end{equation}
In the above expressions, $A_m$ sets the characteristic comoving scale of the perturbation and is chosen 
{
so that the corresponding physical scale is} much larger than the initial cosmological horizon, $R_H\equiv H^{-1}$.

The formation of a PBH is identified through the appearance of a future outer trapping horizon (see Supplemental Material~\cite{SupplementalMaterial}).
Specifically, we monitor the expansions of outgoing and ingoing radial null geodesics, $\Theta_{+}$ and $\Theta_{-}$,
and locate the first radius at which $\Theta_{+}=0$ while $\Theta_{-}<0$.
The corresponding Misner--Sharp mass evaluated at the apparent horizon defines the PBH mass $\MBH$.

For comparison with the PBH literature, we also characterize the initial data by the maximum compaction function, $\mathcal{C}_{\rm max}$, evaluated on our Misner--Sharp slicing (comoving to a fiducial radiation fluid),
\begin{equation}\nonumber
\mathcal{C}(t,A) \equiv \frac{2\,[M(t,A)-M_{\rm b}(t,A)]}{R(t,A)} .
\end{equation}
where $M(t,A)$ is the Misner--Sharp mass and $M_{\rm b}(t,A)$ denotes the corresponding background mass.

\textit{Critical scaling and discrete self-similarity.}
We first determine the collapse threshold for each family of initial data.
In terms of the tuning parameter $p$, the threshold value $p_{\rm th}$ is obtained by bisection,
separating dispersal from PBH formation with absolute precision
$|p-p_{\rm th}|\sim 10^{-8}$.
For the Gaussian and piecewise rational curvature profiles we find
\begin{equation}\nonumber
p_{\rm th}^{\rm G} = 0.089573044, \qquad
p_{\rm th}^{\rm P} = 0.046397240.
\end{equation}
The corresponding threshold values expressed in terms of the maximum of the compaction function are
\begin{equation}\nonumber
\mathcal{C}_{\max,{\rm th}}^{\rm G} = 0.55954879, \qquad
\mathcal{C}_{\max,{\rm th}}^{\rm P} = 0.53554784.
\end{equation}

\begin{figure}[htbp]
\includegraphics[width=0.9\linewidth]{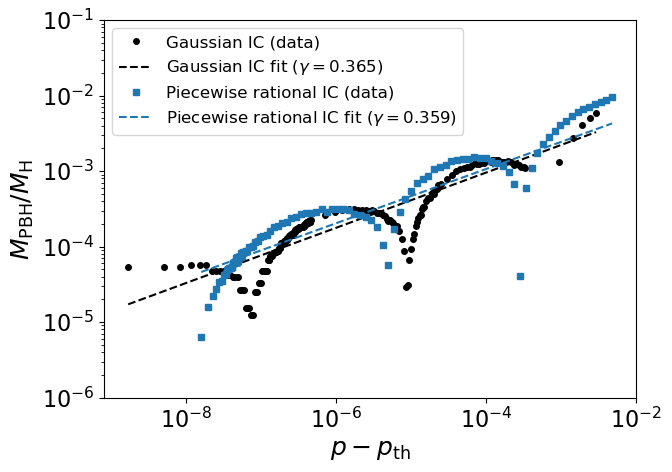}
\caption{\footnotesize{
PBH mass scaling for Gaussian (black) and piecewise rational (blue) initial data,
with the PBH mass normalized by the horizon mass at horizon entry.
Both profiles follow the same power-law envelope with superimposed log-periodic oscillations,
consistent with a common DSS period.
}}
\label{fig:mass_scaling}
\end{figure}

Having determined the collapse threshold, we probe the near-critical regime by tuning $p$
slightly above $p_{\rm th}$ and measuring the resulting PBH masses.
The corresponding mass-scaling relations for both families of initial data are shown in
Fig.~\ref{fig:mass_scaling}.
In both cases the black-hole mass follows a power-law envelope in $(p-p_{\rm th})$
with superimposed log-periodic oscillations characteristic of DSS critical collapse,
consistent with the approximate repetition of suitably rescaled near-critical field
profiles in self-similar variables (see Supplemental Material~\cite{SupplementalMaterial}).
{The DSS modulation makes $\MBH$ non-monotonic in $p$ and produces deep
minima at specific phases of the echoing cycle. Resolving their detailed
behavior would require higher resolution near these phases; whether the minima
remain finite or approach zero is left for future investigation.\footnote{{The masses reported here are formation masses. Post-formation accretion may increase these masses and smooth deep minima in the final PBH mass function, but it does not change their origin in the DSS modulation of the near-critical collapse process.}}} Previous studies of asymptotically flat scalar-field collapse modeled this oscillatory
modulation as sinusoidal
\citep{Rinne:2020asi}.
In our simulations, however, the downward excursions appear sharper than the upward peaks.
While mild systematic deviations from a sinusoidal form can also be observed
in asymptotically flat collapse \citep{Rinne:2020asi}, the substantially
larger number of simulations performed in this work ($\sim 10^3$ for each initial condition)
allows for a more detailed sampling of the near-critical regime.
This extended dataset suggests somewhat more complex oscillatory structure:
a simple sinusoidal model captures the overall modulation but does not fully reproduce
the detailed shape of the residuals (see Supplemental Material~\cite{SupplementalMaterial}).

Following Eq.~\eqref{eq:mass_scaling_with_wiggles}, we extract the critical exponent
$\gamma$ and the logarithmic oscillation period $P_{\ln}$ by fitting the mass-scaling relation.
To determine the oscillation period independently, we perform a Lomb--Scargle analysis
of the residuals after subtracting the power-law envelope (see Supplemental Material~\cite{SupplementalMaterial}).
This procedure yields
\begin{equation}\nonumber
\gamma_{\rm G}=0.365\pm0.017,
\qquad
P_{\ln,{\rm G}}=4.67\pm0.10,
\end{equation}
while for piecewise rational initial data we find
\begin{equation}\nonumber
\gamma_{\rm P}=0.359\pm0.019,
\qquad
P_{\ln,{\rm P}}=4.74\pm0.14.
\end{equation}

{For comparison, in the asymptotically flat collapse of a massless scalar field the DSS critical solution has
$\gamma_{\rm AF} \simeq 0.374$ and $P_{\ln,{\rm AF}} \simeq 4.60$ \citep{PhysRevLett.70.9,Hod:1996az}.
In practice, extracting $\gamma$ and $P_{\ln}$ requires sampling extremely close to criticality. In our numerical simulations—which evolve the collapse on a cosmological background—we were able to reach $(p-p_{\rm th})\sim10^{-8}$, corresponding to a finite dynamically accessible near-critical mass range. {This level of fine-tuning of the initial-data parameter is comparable to that achieved in the original asymptotically flat studies of Choptuik as well as in many subsequent numerical investigations of critical behaviour. The resolved range therefore allows a meaningful comparison with previous numerical results, although its finite extent does not establish conclusively that the fully asymptotic limit has been reached.} The fitted parameters should therefore be regarded as effective values extracted over the available window. Nevertheless, both families of initial data yield mutually consistent results, and the measured $\gamma$ and $P_{\ln}$ remain within $1\sigma$ of the asymptotically-flat benchmark, supporting the interpretation that the local near-critical dynamics approaches the standard DSS {intermediate} attractor.} {Note also that the measured values for $\gamma$ are clearly distinct from the CSS value $\sim 0.94$~\citep{Brady:2002iz}.}

{\textit{Survival of DSS in an expanding universe.}
The persistence of DSS in a cosmological background can be understood from the hierarchy of scales that develops near the black-hole threshold. Although the Friedmann–Lemaître–Robertson–Walker geometry {(FLRW)} introduces the Hubble radius $H^{-1}$ as a natural large-scale reference, the region that undergoes critical collapse becomes progressively smaller as the threshold is approached. In particular, the characteristic size of the collapsing core scales as $(p-p_{\rm th})^\gamma$, and can therefore be made arbitrarily small compared to the cosmological horizon.}

{As a consequence, sufficiently close to criticality the dynamics in the central region is governed by local nonlinear gravitational collapse, while the cosmological expansion primarily affects only the large-scale exterior and its matching to the {FLRW} background. The {near-central} evolution thus approaches the same universal DSS critical solution known {in} asymptotically flat spacetime. Our results indicate that discrete self-similarity survives in an expanding universe as a manifestation of local critical dynamics, with cosmological effects entering through the global structure of the spacetime rather than altering the core collapse itself.} 

\textit{Implications for primordial black-hole mass functions.}
The presence of DSS can leave characteristic imprints on the PBH mass function.
The log-periodic modulation in Eq.~\eqref{eq:mass_scaling_with_wiggles}
modifies the mapping between the collapse parameter
and the resulting PBH mass through a modulation of the Jacobian
relating $p$ and $M_{\rm PBH}$.

At a given horizon mass, the PBH mass function may be written as
\begin{equation}
\psi(\ln M_{\rm PBH}) \equiv \frac{dn_{\rm PBH}}{d\ln M_{\rm PBH}}
\propto
\sum_i P(\delta_H)
\left|\frac{d\delta_H}{d\ln M_{\rm PBH}}\right|_{\delta_{H,i}},
\label{eq:pbm_mass_function}
\end{equation}
where the sum runs over all supercritical solutions
$\delta_{H,i} > \delta_{H,\mathrm{th}}$
satisfying the mass-scaling relation for a fixed
$M_{\mathrm{PBH}}$, $\delta_H = C_{\rm max}$ and $P(\delta_H)$ is the distribution of the maximum of the compaction function.
{We work at fixed horizon mass, corresponding to an infinitely narrow distribution of formation scales. Even in this limit, critical scaling maps the continuous distribution of supercritical amplitudes into an extended PBH mass function, producing the minimum width familiar from standard critical-collapse calculations \citep{Green:1999yh,Gow:2020bzo}.}
Using $\delta_H$ as our control parameter,
\begin{equation}\nonumber
\ln \MBH = \gamma \ln(\delta_H-\delta_{\rm H, th})+ f\left[\ln(\delta_H-\delta_{\rm H, th})\right],
  \end{equation}
  so that
  \begin{equation}\nonumber
  \frac{d\ln M_{\rm PBH}}{dx}
  =
  \gamma + f'(x),
  \qquad
  x=\ln(\delta_H-\delta_{\rm H,th}),
  \end{equation}
  and the Jacobian acquires a log-periodic modulation.

We model the oscillatory contribution phenomenologically using a periodic function motivated by the numerical residuals (see Supplemental Material~\cite{SupplementalMaterial} for details),
\begin{equation}
f(x)= C + A\ln\left[\epsilon+
\left|
\cos\left(\pi\frac{x}{P_{\ln}}+\phi\right)
\right|
\right],
\end{equation}
where $P_{\ln}$ is fixed independently from the residual analysis, {$C\simeq 0$ and $A\simeq 0.3$ from a least-square fitting, and $\epsilon$ is a small positive regularisation parameter}.

\begin{figure}[htbp]
\includegraphics[width=0.9\linewidth]{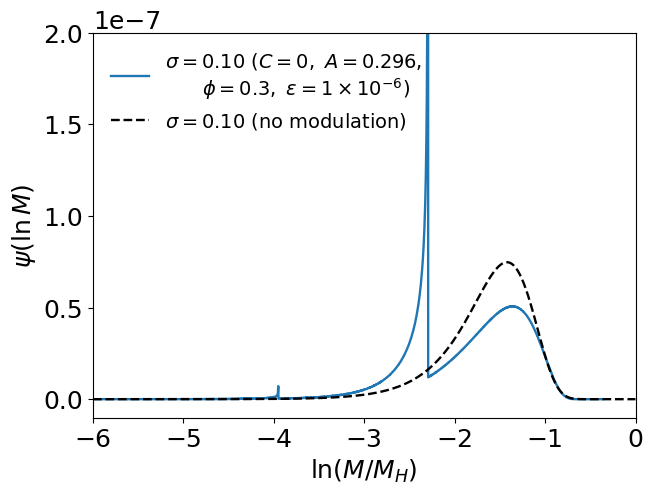}
\caption{\footnotesize{
Illustrative PBH mass functions derived from the critical scaling relation.
The dashed curve corresponds to the case without DSS modulation,
while the solid curve includes a log-periodic modulation of the form
$f(x)$ described in the text.
We assume a Gaussian distribution $P(\delta_H)$.
{The horizon mass is held fixed, so the width of the mass function is generated by critical scaling rather than by a finite-width primordial power spectrum.}
The DSS modulation redistributes statistical weight in mass space,
producing {not only} shifts and distortions of the dominant peak {but also spikes due to the non-monotonicity of $M_{\rm PBH}(\delta_{H})$.}}}
\label{fig:mass_function}
\end{figure}

The DSS-induced modulation redistributes statistical weight in mass space,
generating characteristic shifts and localized enhancements whose spacing
is set by the universal DSS period. In particular, since the mass function involves the inverse of $\gamma+f'(x)$,
the modulation can lead to sharp, resonance-like enhancements
when $\gamma+f'(x)$ becomes small, corresponding to a local compression
of the mapping between $\delta_H$ and $M_{\rm PBH}$.
These features are smoothed by the finite width of $P(\delta_H)$
and by the regularization parameter $\epsilon$, but can still produce
pronounced distortions in the mass spectrum. Figure~\ref{fig:mass_function} illustrates a representative realization of this behavior{, where spikes are generated by the non-monotonicity of $M_{\rm PBH}(\delta_{H})$}.
{{Here $P(\delta_H)$ is a zero-mean Gaussian distribution with standard deviation $\sigma$, and the horizon mass is held fixed. The figure is therefore a diagnostic construction that isolates the DSS imprint of the collapse dynamics, rather than a mass function derived from a finite-width primordial power spectrum. A realistic calculation additionally requires an integration over formation scales. For a narrow primordial spectrum, PBH production is localized around a small range of horizon masses, and the corresponding DSS patterns remain approximately aligned in logarithmic mass, allowing the modulation to survive. For a broad spectrum, contributions from widely separated horizon masses superpose with displaced peaks and troughs and progressively average out. Since the characteristic DSS spacing is $\Delta\ln M_{\rm PBH}\simeq\gamma P_{\ln}\simeq1.7$, integration over several such intervals is expected to strongly suppress the signal. The detailed survival depends on the smoothing prescription, as studied quantitatively in a companion work~\citep{Padilla:2026DSSMassFunction}.}}

\textit{Conclusions.}
In this Letter we have demonstrated that discrete self-similarity (DSS),
originally discovered in the asymptotically flat collapse of a massless scalar field,
also emerges in primordial black-hole (PBH) formation within an expanding cosmological background.
Fully relativistic simulations reveal clear log-periodic modulations in the PBH mass-scaling relation,
providing direct evidence for a DSS critical solution in the cosmological setting.

The critical exponents and logarithmic periods extracted from two distinct families of initial data
are mutually consistent and close to the asymptotically flat benchmarks,
supporting the interpretation that DSS is a local property of critical collapse,
largely insensitive to the global cosmological expansion.
The extended dataset also reveals departures from a simple sinusoidal modulation.
{Although the two initial-data families yield mutually consistent logarithmic periods, differences in the fitted phase and in the detailed manner in which the DSS modulation is sampled suggest that a sufficiently resolved PBH mass pattern could, in principle, retain some information about the primordial perturbation profile. Establishing this connection quantitatively requires a broader survey of initial data.}

The DSS modulation can produce characteristic log-periodic structure in the PBH mass function, suggesting that, when this structure survives the averaging over formation scales, PBHs may probe not only primordial fluctuation amplitudes but also the critical behaviour realized in the early Universe.
{Exact DSS relies on the scale-free massless-scalar system. In realistic kination scenarios, a subdominant potential, radiation component, or scalar-field decay introduces additional scales; if these effects remain negligible over the near-critical region, the DSS solution may persist as an intermediate attractor over a finite range, with the modulation eventually truncated or smoothed.}

\begin{acknowledgments}
{T.H. is grateful to Ilia Musco {and Gabriele Palloni} for fruitful discussion.}
L.E.P. acknowledges support from 
{
JSPS KAKENHI Grant Number~JP25KF0278.}
T.H. acknowledges support from JSPS KAKENHI {Grant Numbers~JP20H05853 and JP24K07027.}
E.M. is supported by an STFC studentship.
D.M. acknowledges support from STFC Grant No.~ST/X000931/1.
\end{acknowledgments}
\bibliographystyle{ieeetr}
\bibliography{biblio}

\newpage

\onecolumngrid
\begin{center}
{\large \textbf{Supplemental Material for:}\\[0.3em]
\textbf{Cosmological discrete self-similarity in primordial black hole formation}}
\end{center}
\twocolumngrid


\section{Evolution equations}

The simulations are performed using the extended Misner--Sharp
formulation described in
Refs.~\citep{Milligan:2025zbu,Padilla:2025bkv}.
For completeness we summarize here the system of equations
solved numerically.

The spacetime metric is written as
\begin{equation}\label{eq:MSmetric}
ds^2 = -e^{2\phi}dt^2 + e^{\lambda}dA^2 + R^2d\Omega^2 ,
\end{equation}
where the metric functions $\phi$, $\lambda$, and the areal radius $R$
depend on the radial coordinate $A$ and time $t$.

The matter sector consists of a perfect fluid and a scalar field,
with energy--momentum tensor
\begin{eqnarray}
T_{\mu\nu} &=& (\rho_{\rm pf} + P_{\rm pf})u_{\mu}u_{\nu}
+ P_{\rm pf} g_{\mu\nu} \nonumber \\
&& - g_{\mu\nu}\left(\frac12\partial_{\sigma}\psi\partial^{\sigma}\psi
+ V(\psi)\right)
+ \partial_{\mu}\psi\partial_{\nu}\psi .
\end{eqnarray}
The coordinate system is chosen to be comoving with the perfect fluid,
\begin{equation}
u^t = e^{-\phi}, \qquad u^i = 0 \quad (i=A,\theta,\phi).
\end{equation}

The background cosmology is assumed to be a flat Friedmann-Lema\^{i}tre-Robertson-Walker spacetime
with scale factor $a(t)$ and Hubble parameter $H=\dot a/a$.
We denote by $R_H = H^{-1}$ the Hubble radius at the initial time $t_i$ and by
$\rho_{\rm b}$ the background total energy density.

Following Refs.~\citep{Milligan:2025zbu,Padilla:2025bkv},
we evolve dimensionless variables denoted by tildes.
These are defined through
\begin{subequations}
\begin{equation}
R = aA\tilde{R},
\end{equation}
\begin{equation}
\rho_{\rm pf} = \rho_{\rm b}\tilde{\rho}_{\rm pf},
\qquad
\rho_\psi = \rho_{\rm b}\tilde{\rho}_\psi,
\end{equation}
\begin{equation}
P_{\rm pf} = \rho_{\rm b}\tilde{P}_{\rm pf},
\qquad
P_{\psi} = \rho_{\rm b}\tilde{P}_{\psi},
\end{equation}
\begin{equation}
m_{\rm pf} = \frac{4\pi}{3}\rho_{\rm b}R^3\tilde{m}_{\rm pf},
\qquad
m_{\psi} = \frac{4\pi}{3}\rho_{\rm b}R^3\tilde{m}_{\psi},
\end{equation}
\begin{equation}
U = HR\tilde{U},
\end{equation}
\begin{equation}
\psi = R_H\sqrt{\rho_{\rm b}}\,\tilde{\psi},
\qquad
\chi = \sqrt{\rho_{\rm b}}\,\tilde{\chi},
\qquad
\Pi = \sqrt{\rho_{\rm b}}\,\tilde{\Pi},
\end{equation}
\end{subequations}
Auxiliary variables are defined as
\begin{equation}
U = e^{-\phi}\dot{R},
\qquad
\Gamma = e^{-\lambda/2}R',
\end{equation}
\begin{equation}
\chi \equiv \psi',
\qquad
\Pi \equiv e^{-\phi}\dot{\psi}.
\end{equation}

The evolution equations are written in terms of the logarithmic time
variable $\xi=\ln t$.
The full system of equations solved numerically is
\begin{widetext}
\begin{subequations}
\begin{equation}
\partial_\xi\tilde{\psi} =
\alpha e^{\phi+\xi}\tilde{\Pi} + \tilde{\psi},
\end{equation}

\begin{equation}
\partial_\xi\tilde{\chi} =
\alpha e^{\phi+\xi}
(\tilde{\Pi}' + \phi'\tilde{\Pi})
+ \tilde{\chi},
\end{equation}

\begin{equation}
\partial_\xi\tilde{\Pi}
=
\tilde{\Pi}
+\alpha e^{\phi}\bigg\lbrace
e^{\xi-\lambda}
\left[
\left(
\frac{2(\bar{A}\tilde{R})'}{\bar{A}\tilde{R}}
+ \phi'
-\frac{\lambda'}{2}
\right)\tilde{\chi}
+ \tilde{\chi}'
\right]
-
\left[
\left(
2\tilde{U}
+
\frac{(\bar{A}\tilde{R}\tilde{U})'
+\frac{3}{2}e^{-\xi}\bar{A}\tilde{R}\tilde{\chi}\tilde{\Pi}}
{(\bar{A}\tilde{R})'}
\right)\tilde{\Pi}
+ e^{\xi}\tilde{V}_{,\tilde{\psi}}
\right]
\bigg\rbrace,
\end{equation}

\begin{equation}
\partial_\xi\tilde{R} =
\alpha\tilde{R}(\tilde{U}e^{\phi}-1),
\end{equation}

\begin{equation}
\partial_\xi\tilde{\rho}_{\rm pf}
=
2\tilde{\rho}_{\rm pf}
-
\alpha e^\phi
\left[
2\tilde U
+
\frac{(\bar{A}\tilde{R}\tilde{U})'
+\frac{3}{2}e^{-\xi}\bar{A}\tilde{R}\tilde{\chi}\tilde{\Pi}}
{(\bar{A}\tilde{R})'}
\right]
(\tilde{\rho}_{\rm pf}+\tilde{P}_{\rm pf}),
\end{equation}

\begin{equation}
\partial_\xi\tilde{U}=
\tilde{U}
-
\alpha e^{\phi}
\left[
\frac{e^{2\xi-\lambda}(\bar{A}\tilde{R})'P'_{\rm pf}}
{(\bar{A}\tilde{R})(\tilde{\rho}_{\rm pf}+\tilde{P}_{\rm pf})}
+
\frac{3}{2}(\tilde{U}^2+\tilde{P}_{\rm T})
+
\frac{e^{2(1-\alpha)\xi}-\bar{\Gamma}^2}
{2(\bar{A}\tilde{R})^2}
\right],
\end{equation}

\begin{equation}
\partial_\xi\lambda =
\alpha e^\phi
\left(
\frac{2(\bar{A}\tilde{R}\tilde{U})'
+3e^{-\xi}\bar{A}\tilde{R}\tilde{\chi}\tilde{\Pi}}
{(\bar{A}\tilde{R})'}
\right),
\end{equation}

\begin{equation}
\phi' =
-\frac{\tilde{P}'_{\rm pf}}
{\tilde{\rho}_{\rm pf}+\tilde{P}_{\rm pf}},
\end{equation}

\begin{equation}\label{eq:HC0}
\bar{\Gamma}^2 = \frac{\Gamma^2}{a^2H^2R_H^2} =
e^{2(1-\alpha)\xi}
+
\bar{A}^2\tilde{R}^2(\tilde{U}^2-\tilde{m}_{\rm T})
+
\frac{3e^{-\xi}}{\bar{A}\tilde{R}}\mathcal{I}.
\end{equation}
\end{subequations}
\end{widetext}

where
\begin{equation}
\mathcal{I} =
\int_0^{\bar{A}_0}
\tilde{U}\bar{A}^3\tilde{R}^3\tilde{\chi}\tilde{\Pi}
\, d\bar{A}.
\end{equation}
In the above equations, primes denote derivatives with respect to the radial coordinate $\bar{A} = A/R_H$. We have also introduced the parameter {$\alpha = {2}/[{3(1+\omega)]}$}, where $\omega$ is the mean equation-of-state parameter of the background universe. 
\section{Misner--Sharp mass and horizon identification}

The PBH mass is defined using the Misner--Sharp mass,
which in spherical symmetry is given by

\begin{equation}\label{eq:MSmass}
M_{\rm MS} = \frac{R}{2}\left(1 - \nabla^a R \nabla_a R \right),
\end{equation}
where $R$ is the areal radius.

The expansions of outgoing and ingoing radial null geodesics are

\begin{equation}
\Theta_\pm =
\frac{2}{\tilde{R}}
\left(
e^{-\tilde{\phi}}\dot{\tilde{R}}
\pm
e^{-\tilde{\lambda}/2}\tilde{R}'
\right).
\end{equation}
A future outer trapping horizon is identified by the conditions

\begin{equation}
\Theta_+ = 0,
\qquad
\Theta_- < 0 .
\end{equation}
In practice the horizon position is obtained by interpolating
between neighboring grid points where the sign of $\Theta_+$
changes.

The PBH mass is then defined as the Misner--Sharp mass
evaluated at {the first appearance of} the apparent horizon,

\begin{equation}
M_{\rm PBH} = M_{\rm MS}(t_{\rm AH},A_{\rm AH}).
\end{equation}

\section{Near-critical self-similar profiles}

A characteristic feature of discretely self-similar (DSS) critical
collapse is the emergence of an intermediate attractor in which
suitably rescaled dynamical variables repeat periodically in
logarithmic time. To test whether the near-threshold dynamics in our
cosmological simulations approaches such an attractor, we analyze the
evolution of selected variables in self-similar coordinates.

In the vicinity of the collapse threshold the characteristic length
scale of the collapsing region shrinks as the critical solution is
approached. The natural time variable describing this evolution is
the proper time measured at the origin, $\tau$. Following the standard
description of DSS collapse \citep{PhysRevLett.70.9,PhysRevD.55.695},
we introduce the logarithmic self-similar time coordinate
\begin{equation}
T = -\ln(\tau_* - \tau),
\end{equation}
where $\tau_*$ denotes the accumulation time of the critical solution.
In terms of this variable, successive ``echoes'' of the DSS solution
are expected to appear periodically in $T$, with a fundamental period
$\Delta$.

To visualize the approach to the self-similar attractor we examine
profiles of the physical energy density {$\rho_{\psi} = -(\nabla_\mu \psi \nabla^\mu\psi)/2 $} and of the scalar field
momentum $\Pi$. Because the energy density is quadratic in the scalar-field variables,
its profile repeats with half the fundamental DSS period, i.e.
$\Delta/2$, rather than $\Delta$. In contrast, the scalar-field
momentum $\Pi$, being linear in the field variables, changes sign
under each half echo and therefore only repeats after a full period
$\Delta$. This leads to an apparent doubling of the echoing frequency
in the quadratic observables shown below.

The echoing period $\Delta$ is extracted from the separation of
successive peaks at the center in the self-similar energy density profiles. Since consecutive peaks
are separated by $\Delta/2$, we estimate the fundamental period from
\begin{equation}
\Delta_i = 2(T_{i+1} - T_i),
\end{equation}
where $T_i$ denotes the self-similar time at which a peak in the energy density occurs. We then define
\begin{equation}
\Delta = \langle \Delta_i \rangle,
\end{equation}
where the average is taken over the set of identified peaks. 

From our simulations we obtain a value of $\Delta \simeq 3.36$, 
consistent within $1\sigma$ with that inferred from the oscillatory 
modulation of the mass scaling relation discussed in the main text.
This value is slightly smaller than the canonical asymptotically flat
result of Choptuik, which may be attributed to the limited number of
resolved echoing cycles in our simulations. In practice, accurately
extracting $\Delta$ requires following the evolution arbitrarily close
to the critical solution, which becomes increasingly challenging as
the characteristic length scale shrinks.

We therefore interpret the value reported here as an effective estimate
of the echoing period. Its proximity to the asymptotically flat result
provides further evidence that the near-critical dynamics is governed
by a DSS-like intermediate attractor. A more precise determination of
$\Delta$, requiring higher resolution and longer evolutions within the
self-similar regime, is left for future work.

\begin{figure}[h]
\includegraphics[width=0.9\linewidth]{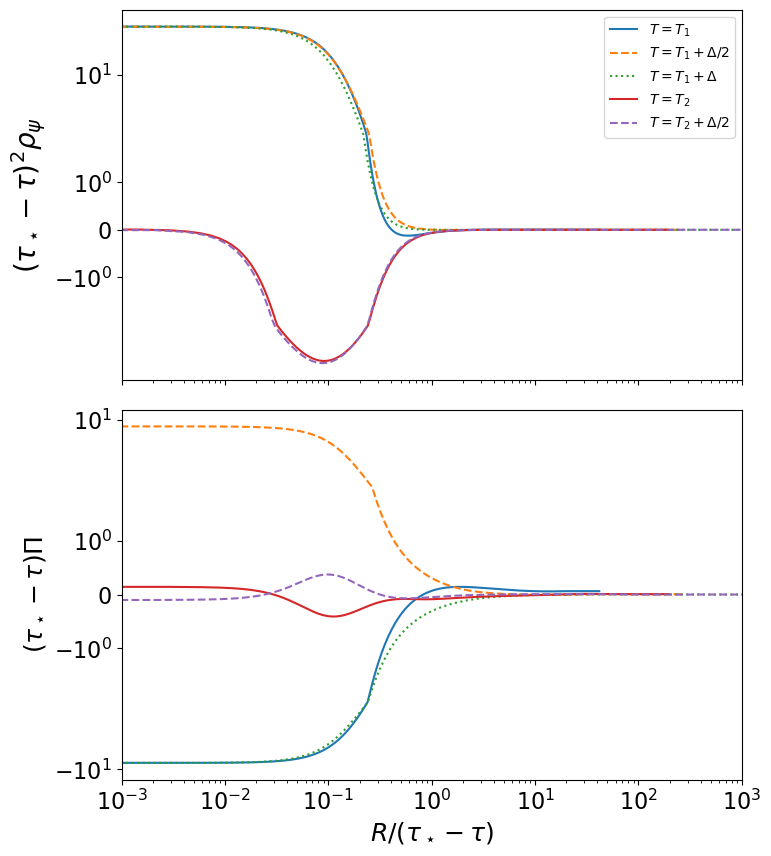}
\caption{\footnotesize{
Near-critical self-similar profiles for a representative evolution.
Profiles of the physical density $\rho_\psi$ (top) and the scalar field
momentum $\Pi$ (bottom) are shown at successive values of the
self-similar time coordinate $T$. The approximate repetition of the
profiles reflects the echoing behavior characteristic of discretely
self-similar collapse. In this example, one profile captures a full
echoing cycle, while a later one shows only a partial cycle due to the
departure from the self-similar attractor.}}
\label{fig:selfsimilar_profiles}
\end{figure}

Figure~\ref{fig:selfsimilar_profiles} shows representative profiles of
{$\rho_{\psi}$} and $\Pi$ for a near-critical subcritical evolution,
evaluated at successive values of the self-similar time coordinate $T$.
As the collapse approaches the threshold, the central region exhibits
repeated episodes of contraction in which the profiles approximately
reproduce the same shape after a logarithmic shift in time and scale,
with spacing $\Delta/2$ in $T$ for the energy density and $\Delta$ for the case of the scalar field momentum $\Pi$, consistent with the presence of a DSS
intermediate attractor governing the near-critical dynamics.

In practice, we observe that while one of the profiles captures an
approximately complete echoing cycle, later profiles only exhibit
a partial repetition. This reflects the fact that, after approaching
the critical solution, the evolution eventually departs from the
self-similar attractor, preventing the full echoing pattern from being
resolved at later times.

We have verified that the same echoing structure and period are
obtained for all initial profiles considered in this work. For clarity, we present here only the case of Gaussian initial
conditions, which provides a representative example of the behaviour
observed across the different initial profiles considered in this work.

\section{Residual oscillations}

\begin{figure}[h]
\includegraphics[width=0.9\linewidth]{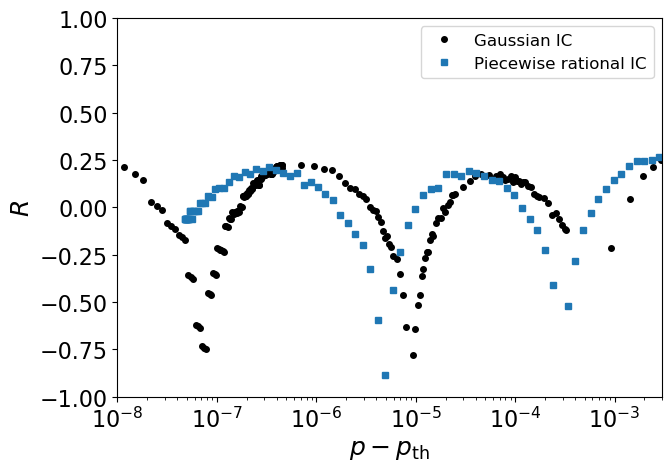}
\caption{\footnotesize{
Residuals of the PBH mass scaling relation after subtraction of the
power-law envelope.
The oscillatory structure reflects the log-periodic modulation associated
with discrete self-similarity.
}}
\label{fig:residuals}
\end{figure}

The PBH mass scaling relation is modeled as

\begin{equation}
\ln M_{\rm PBH}
=
\gamma \ln(p-p_{\rm th})
+
f[\ln(p-p_{\rm th})].
\end{equation}
To determine the oscillatory component we subtract the best-fit
power-law envelope,

\begin{equation}\label{eq:sl}
\ln M_{\rm PBH} \approx
\gamma \ln(p-p_{\rm th}) + \ln M_0.
\end{equation}
The residuals are defined then as
\begin{equation}\label{eq:residual}
R =
\ln M_{\rm PBH}
-
\left(
\gamma \ln(p-p_{\rm th}) + \ln M_0
\right).
\end{equation}
This definition isolates the oscillatory component associated
with discrete self-similarity. In Fig.~\ref{fig:residuals} we show the residual for each of our initial profiles.

\section{Lomb--Scargle analysis}

\begin{figure}[h]
\includegraphics[width=0.9\linewidth]{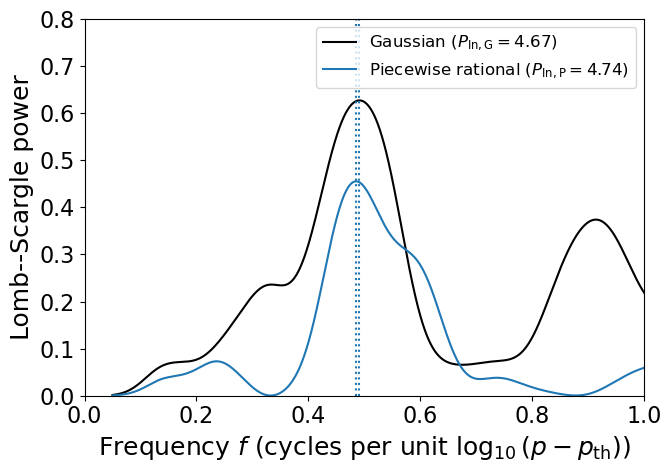}
\caption{\footnotesize{
Lomb--Scargle periodogram of the residual signal.
The dominant peak determines the oscillation period of the
log-periodic modulation.
}}
\label{fig:lomb}
\end{figure}

Because the sampling in $\ln(p-p_{\rm th})$ is not uniform,
we determine the oscillation period using a Lomb--Scargle
periodogram applied to the residuals. This method is well suited for unevenly sampled data. The dominant peak of the Lomb--Scargl spectrum shown in Fig.~\ref{fig:lomb} yields the
oscillation period $P_{\ln}$.

\section{Comparison with a simple sinusoidal fit}

As a first approximation, it is natural to model the residual signal
defined in Eq.~(\ref{eq:residual}) with a simple sinusoidal template,
\begin{equation}
R_{\rm sin}(x)= C_0 + C_1 \cos\!\left(\frac{2\pi}{P_{\ln}}x+\phi\right),
\end{equation}
where $x=\ln(p-p_{\rm th})$, $C_0$ and $C_1$ are fitting constants, $\phi$ is a phase shift,
and the period $P_{\ln}$ is fixed from the Lomb--Scargle analysis.

This simple form captures the overall oscillatory trend of the numerical
data reasonably well and provides a useful first description of the
log-periodic modulation. However, it does not fully reproduce the
detailed morphology of the residuals. In particular, the numerical
signal exhibits slightly sharper downward excursions and broader maxima
than those predicted by a purely sinusoidal template.

This difference is not dramatic, and the sinusoidal fit remains useful
as a baseline description. Nevertheless, the larger number of
simulations included in this work allows for a more detailed sampling
of the near-critical regime, which suggests that the oscillatory
structure is somewhat more intricate than a simple harmonic form.

\begin{figure}[h]
\includegraphics[width=0.9\linewidth]{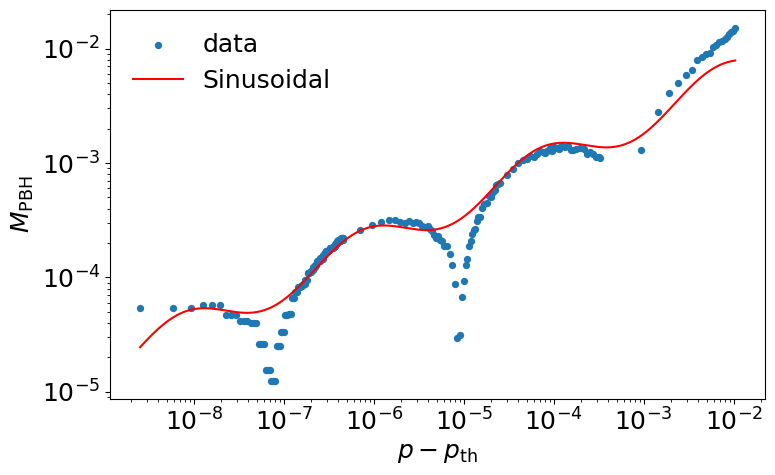}
\caption{\footnotesize{
PBH mass scaling relation for the Gaussian initial profile. The points show the numerical data, while the solid curve shows the combined fit consisting of a linear power-law trend plus a sinusoidal modulation. Although this model captures the overall log-periodic behavior, it does not fully reproduce the detailed structure of the oscillations.
}}
\label{fig:sin_full}
\end{figure}

Figure~\ref{fig:sin_full} shows the numerical data of the initial Gaussian profile together with the
best-fit power-law envelope and the corresponding sinusoidally modulated
model.

\section{Phenomenological fit of the oscillatory modulation}

Although the simple sinusoidal template discussed above captures the
overall oscillatory behavior, it does not fully account for the detailed
shape of the residual signal as a function of
$x=\ln(p-p_{\rm th})$.
In particular, the oscillations exhibit slightly sharper minima and
broader maxima, suggesting that a more flexible periodic template may
provide a better effective description of the numerical data.

To model this behaviour, we introduce a phenomenological fitting
function of the form
\begin{equation}
f(x)= C + A\,\ln\!\left[\epsilon+
\left|
\cos\!\left(\pi\frac{x}{P_{\ln}}+\phi\right)
\right|
\right],
\label{eq:phenom_fit}
\end{equation}
where $C$ is a constant vertical offset, $A$ controls the amplitude of
the modulation, $\phi$ is a phase shift, and $\epsilon>0$ is a small
regularization parameter introduced to keep the logarithm finite near
the minima of the oscillation.

The argument of the cosine is written explicitly in terms of the
logarithmic period $P_{\ln}$, so that the function is periodic under
$x \to x + P_{\ln}$. The period $P_{\ln}$ is not treated as a free
parameter but is fixed from the Lomb--Scargle analysis of the residuals.

\begin{figure}[h]
\includegraphics[width=0.9\linewidth]{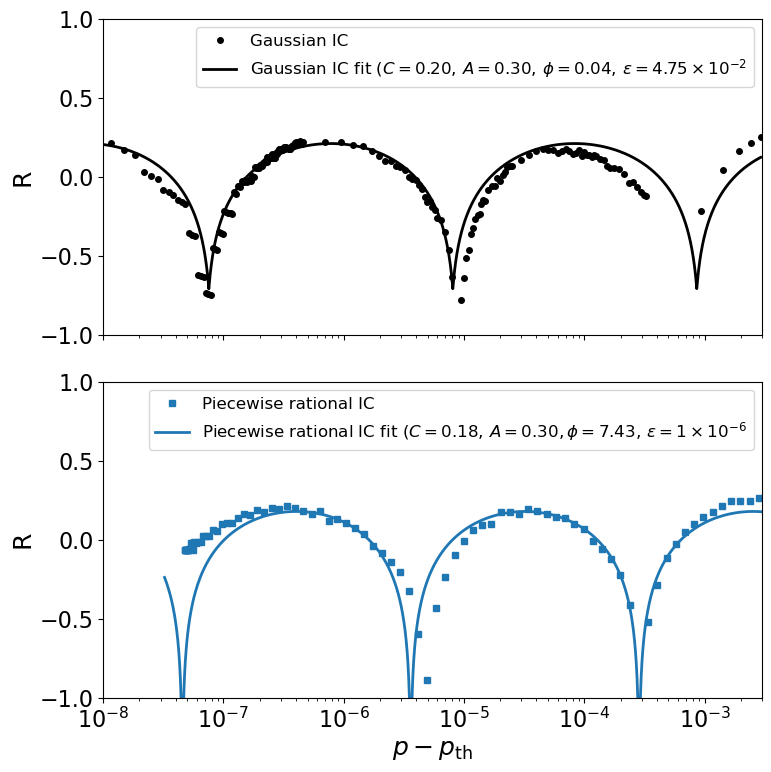}
\caption{\footnotesize{
Phenomenological fit of the oscillatory residual signal using
Eq.~(\ref{eq:phenom_fit}).
The logarithmically distorted periodic template reproduces the
non-sinusoidal structure of the residuals, in particular the sharper
minima and broader maxima seen in the simulations.
}}
\label{fig:phenom_fit}
\end{figure}

This form is not intended to represent a fundamental prediction for the
critical solution, but rather to provide a simple phenomenological
description of the observed residual pattern.
The logarithm enhances the sharpness of the minima, while the absolute
value produces a sequence of cusp-like troughs and broader maxima, in
better qualitative agreement with the numerical residuals than a pure
sinusoid.
The remaining parameters $(C,A,\phi,\epsilon)$ are determined through a
nonlinear least-square fit. Figure~\ref{fig:phenom_fit} shows the residual
signal together with the best-fit phenomenological model. Although this
specific fitting function is not unique, and other functional forms may
also reproduce the observed oscillations, the overall shape of the
modulation appears to be robust across our datasets. In this sense, what
may be universal is not the particular fitting ansatz itself, but the
general oscillatory structure it is meant to capture. Correspondingly, the
fitted parameters should be interpreted mainly as effective quantities
characterizing this structure rather than as fundamental predictions of the
critical solution. We note, however, that the fitted values are broadly
consistent between the two families of initial data, with the exception of
$\epsilon$. This parameter is more difficult to determine reliably, since
it is especially sensitive to the simulations whose masses lie near the
local minima of the oscillatory pattern, where a higher degree of numerical
control is required.

\section{Numerical convergence}

The numerical implementation employed in this work is based on the
extended Misner--Sharp formulation developed in
Refs.~\citep{Milligan:2025zbu,Padilla:2025bkv}. The code has been
extensively tested in previous studies across different scalar-field
cosmologies and collapse scenarios. Here we restrict ourselves to two
consistency checks relevant for the present simulations.

First, we measure the convergence of the numerical scheme by monitoring
the Hamiltonian constraint evaluated during the evolution.
Second, we evaluate the Hamiltonian constraint for all simulations at
the time when the PBH mass is extracted, providing a global diagnostic
of the constraint violation across the parameter space explored in the
main analysis.

In the grid interior, spatial derivatives are evaluated using
fourth-order finite-difference stencils, while time integration is
performed with the adaptive Runge--Kutta solver \texttt{DOP853} from
\texttt{solve\_ivp}. Nevertheless, the observed convergence of monitored
constraint residuals is not determined solely by the formal order of
these interior evolution operators. In practice, boundary closures,
regularity conditions at the origin, and auxiliary interpolation and
integration procedures affect the effective convergence order of
derived quantities.

To assess the numerical accuracy of the implementation we perform
simulations at resolutions $N = 700$, $1400$, and $2800$ and monitor
the Hamiltonian constraint.

The Hamiltonian constraint is evaluated by rewriting the Einstein
constraint equation \eqref{eq:HC0} into an equivalent form,
\begin{eqnarray}\label{eq:HC}
   {\mathscr{H}} &=& \bar A^2 \tilde R^2 \tilde U\left[{(\bar A\tilde{R}\tilde{U})^{\prime} + \frac{3}{2}e^{-\xi}\Bar{A}\tilde{R}\tilde{\chi}\tilde{\Pi}}\right] \nonumber \\
    && - {(\bar A\tilde R)^{\prime}}\left[\frac{\bar A\tilde R\bar \Gamma'}{e^{\lambda/2-\xi}}+\frac{3}{2}\bar A^2\tilde R^2\tilde \rho_{\rm T} \right. \\
    &&\left.+ \frac{\bar \Gamma^2 -(\bar A\tilde R\tilde U)^2 - e^{2(1-\alpha)\xi}}{2}\right].\nonumber 
\end{eqnarray}
which should be equal to zero in the continuum limit. This form avoids the need to evaluate the integral in Eq.~(\ref{eq:HC0})
at every time step, which would be computationally expensive. In the numerical
implementation we monitor the normalized residual
\begin{equation}
    \mathcal{H} = \frac{|\mathscr{H}|}{\sum_i |T_i|},
\end{equation}
where \( T_i \) denotes each of the individual terms contributing to
\( \mathscr{H} \), i.e., \( \mathscr{H} = \sum_i T_i \).

\begin{figure}[h]
\includegraphics[width=0.9\linewidth]{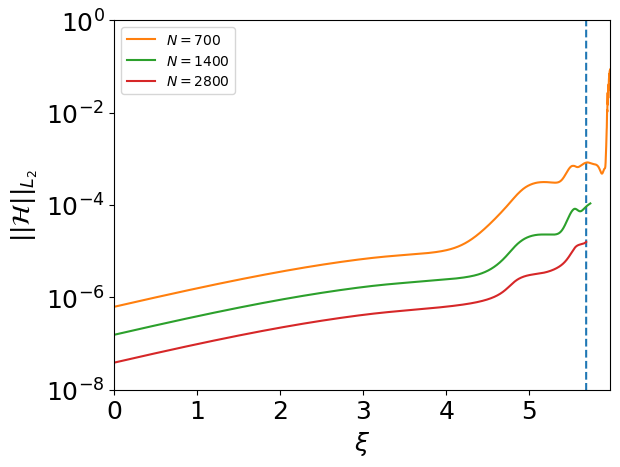}
\caption{\footnotesize{
Convergence test of the $L_2$ norm of the Hamiltonian-constraint
residual for three grid resolutions.
The violation decreases approximately as $\Delta x^2$, indicating
effective second-order convergence of this diagnostic.
The dashed blue line marks the time of apparent-horizon formation in
the simulation. The PBH mass reported in the main text is measured at
this moment from the Misner--Sharp mass evaluated at the apparent
horizon.
}}
\label{fig:convergence}
\end{figure}

Figure~\ref{fig:convergence} shows the convergence test obtained from
the $L_2$ norm of the Hamiltonian constraint residual for three grid
resolutions. The violation decreases approximately as
{$(\Delta x)^2$}, indicating effective second-order convergence for this
diagnostic in the full numerical scheme.

In addition to the resolution study shown above, we also monitor the
Hamiltonian constraint across the full set of simulations used in the
mass-scaling analysis.
For each run we evaluate the $L_2$ norm of the Hamiltonian constraint
at the time when the PBH mass is measured, namely at the formation of
the apparent horizon.

\begin{figure}[h]
\includegraphics[width=0.9\linewidth]{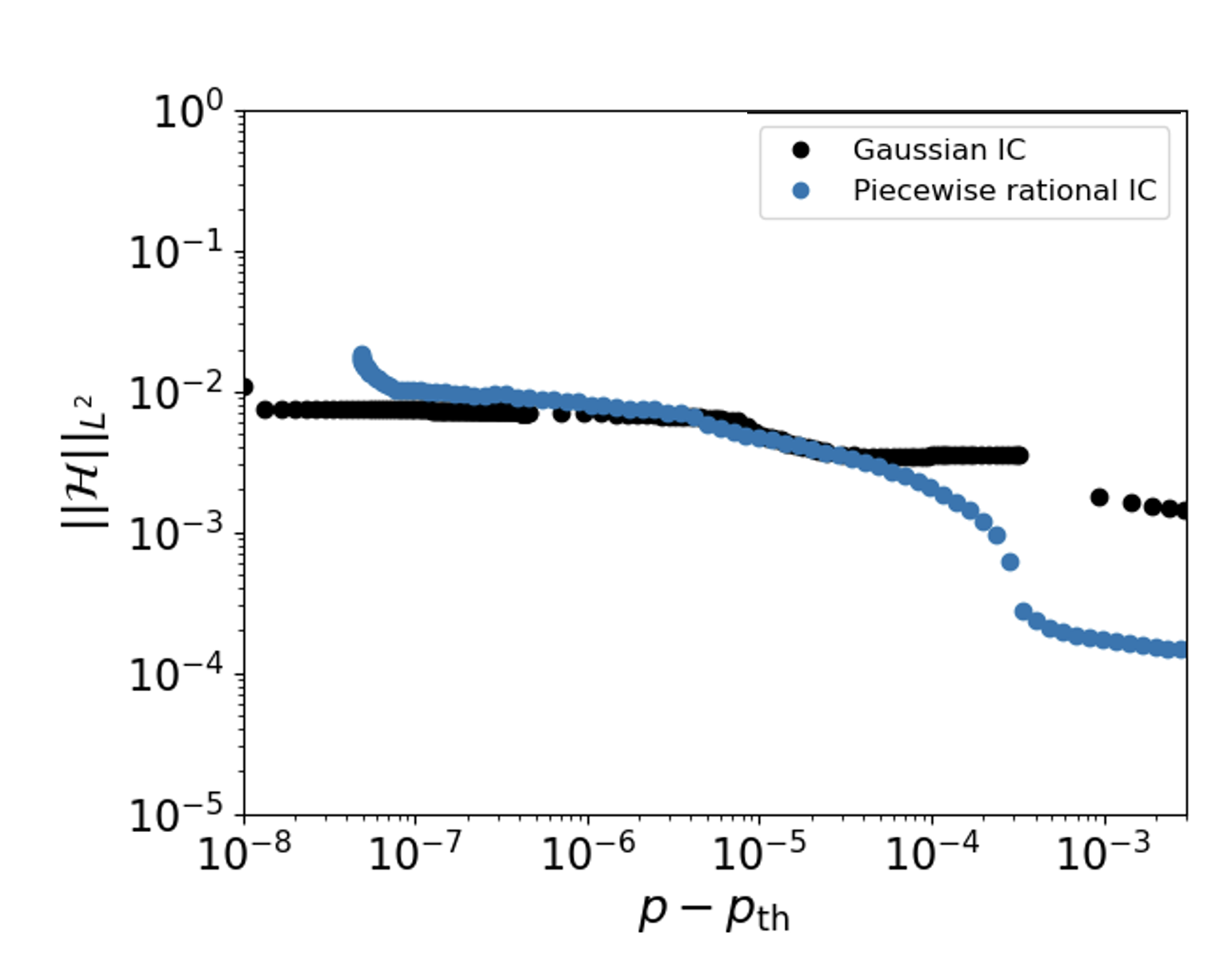}
\caption{\footnotesize{
$L_2$ norm of the Hamiltonian constraint evaluated at the time of
apparent-horizon formation for all simulations used in the PBH mass
scaling analysis.
Each point corresponds to one simulation in the parameter scan.
The small magnitude of the constraint residual indicates that the
PBH masses are extracted from numerically consistent evolutions.
}}
\label{fig:HC_global}
\end{figure}

Figure~\ref{fig:HC_global} shows the resulting constraint violation for
all simulations considered in this work.
The residual remains small for the entire dataset, confirming that the
PBH masses reported in the main text are extracted from numerically
well-resolved evolutions.
\end{document}